\documentstyle[epsfig]{aipproc}

\newcommand\prd[3]{Phys. Rev. D {\bf #1}, #2 (#3)}
\newcommand\prl[3]{Phys. Rev. Lett. {\bf #1}, #2 (#3)}
\newcommand\apj[3]{Astrophys. J. {\bf #1}, #2 (#3)}
\newcommand\apjs[3]{Astrophys. J. Suppl. {\bf #1}, #2 (#3)}

\begin{document}

\title{The Innermost Stable Circular Orbit in Compact Binaries}

\author{Thomas W. Baumgarte}

\address{Department of Physics, 
        University of Illinois at Urbana-Champaign, Urbana, Il 61801}

\maketitle

\begin{abstract}
Newtonian point mass binaries can be brought into arbitrarily close
circular orbits.  Neutron stars and black holes, however, are
extended, relativistic objects.  Both finite size and relativistic
effects make very close orbits unstable, so that there exists an
innermost stable circular orbit (ISCO).  We illustrate the physics of
the ISCO in a simple model problem, and review different techniques
which have been employed to locate the ISCO in black hole and neutron
star binaries.  We discuss different assumptions and approximations,
and speculate on how differences in the results may be explained and
resolved.
\end{abstract}

\section*{Introduction}

Compact binaries, containing black holes or neutron stars, are among
the most promising sources for the new generation of gravitational
wave detectors.  TAMA has already started taking data, and LIGO, VIRGO
and GEO will soon become operational (see, e.g.,~\cite{b00b}).  With
the advent of gravitational wave astronomy arises a need for
theoretical templates of gravitational waveforms, which are required
for the identification and interpretation of signals in the noisy
output of the detectors.  

Compact binaries emit gravitational radiation, and therefore loose
energy and slowly spiral towards each other.  Because of the
circularizing effects of gravitational radiation damping, it is
reasonable to assume the orbits of close binaries to be
quasi-circular.  The slow and adiabatic inspiral continues until the
binary reaches the innermost stable circular orbit (ISCO), at which
the orbits become unstable.  At that point, the stars start to plunge
towards each other, and coalesce and merge after a dynamical
timescale.  The ISCO leaves a characteristic signature in the
gravitational wave signal of a binary inspiral, and therefore provides
an important piece of information for the construction of
gravitational wave templates.

In this article, we will explain in a simple point-mass model problem
how both tidal and relativistic effects make very close orbits
unstable, giving rise to the ISCO.  We will then review attempts
to determine the ISCO for binary black holes, which, for stellar-mass
black holes, is likely to fall into the frequency range of LIGO.  In
particular, we will compare results from post-Newtonian (PN) and
numerical calculations.  We will also discuss some of the qualitative
results for binary neutron stars, and will summarize some of the more
recent results from relativistic simulations.

\section*{The ISCO in a simple model problem}

Most commonly, the ISCO is located with the help of turning-point
methods.  We motivate this method by applying it to Newtonian
point-masses, and introduce tidal and relativistic correction terms to
illustrate their effect.  Rigorous justifications for
turning-point methods have been developed, for example, 
in~\cite{lrs93b,lrs97,bcsst98a}.

\subsection*{Newtonian point-masses}

The conditions for a circular orbit can be derived very easily from
the Hamiltonian formalism.  Consider, for example, a Newtonian point
mass binary, for which the energy is the sum of the kinetic and
potential energy
\begin{equation}
E = \frac{1}{2} \mu \, \dot {\bf r}^2 - \frac{\mu M}{r}
	= \frac{1}{2} \mu (\dot r^2 + r^2 \Omega^2) - \frac{\mu M}{r}.
\end{equation}
Here $M \equiv m_1 + m_2$ is the total mass, $\mu \equiv m_1 m_2/M$
the reduced mass, ${\bf r}$ the separation vector, and $\Omega = \dot
\varphi$ the angular velocity.  Rewriting the energy in terms of the
conjugate momenta $P = \mu \dot r$ and $J = \mu r^2 \Omega$ yields the
Hamiltonian
\begin{equation} \label{E1}
E = H = \frac{1}{2}\,\frac{P^2}{\mu} 
	+ \frac{1}{2}\, \frac{J^2}{\mu r^2}
	- \frac{\mu M}{r}.
\end{equation}
The four independent variables $r$, $P$, $\varphi$ and $J$ now satisfy
the Hamiltonian equations
\begin{equation} \label{heom}
\begin{array}{rclrcl}
\dot r & = & \displaystyle \frac{\partial H}{\partial P}
~~~~~~~~&~~~~~~~~
\dot P & = & \displaystyle - \frac{\partial H}{\partial r} \\[3mm]
\Omega ~ = ~ \dot \varphi & = & \displaystyle \frac{\partial H}{\partial J} &
\dot J & = &  \displaystyle - \frac{\partial H}{\partial \varphi} ~ = ~ 0. 
\end{array}
\end{equation}
The last equation shows that the angular momentum is conserved, since
the Hamiltonian is cyclic in $\varphi$.  To construct a circular
orbit, we obviously need $\dot r = 0$, which implies $P = 0$, and also
$\dot P = 0$, so that
\begin{equation} \label{dEdr}
\left. \frac{\partial H}{\partial r} \right|_J = 
\left. \frac{\partial E}{\partial r} \right|_J = 0.
\end{equation}
A circular orbit hence corresponds to an extremum of the energy at
constant angular momentum.  The orbital frequency of a circular orbit
can be determined from
\begin{equation} \label{dEdJ}
\Omega = \displaystyle \frac{\partial H}{\partial J} 
	= \displaystyle \frac{\partial E}{\partial J}.
\end{equation}
The last two equations are crucial for the construction of the ISCO.

Returning to the example of a Newtonian point mass binary, we can
construct circular orbits by inserting the energy~(\ref{E1}) with
$P=0$ into eqs.~(\ref{dEdr}) and~(\ref{dEdJ}).  Eq.~(\ref{dEdr}) yields
the virial relation
\begin{equation} \label{virial}
\frac{J^2}{\mu r^2} = \frac{\mu M}{r},
\end{equation}
so that the equilibrium energy of a circular orbit becomes
\begin{equation} \label{Eeq}
E_{\rm eq} = - \frac{1}{2}\,\frac{\mu M}{r}.
\end{equation}
The orbital frequency of these orbits can be found from eq.~(\ref{dEdJ})
and satisfies, as expected, Kepler's third law
\begin{equation}
\Omega = \frac{\partial E}{\partial J} = 
	\frac{J}{\mu r^2} = \sqrt{\frac{M}{r^3}},
\end{equation}
where we have used eq.~(\ref{virial}) in the last equality.  

\begin{figure}[t] 
\centerline{\epsfig{file=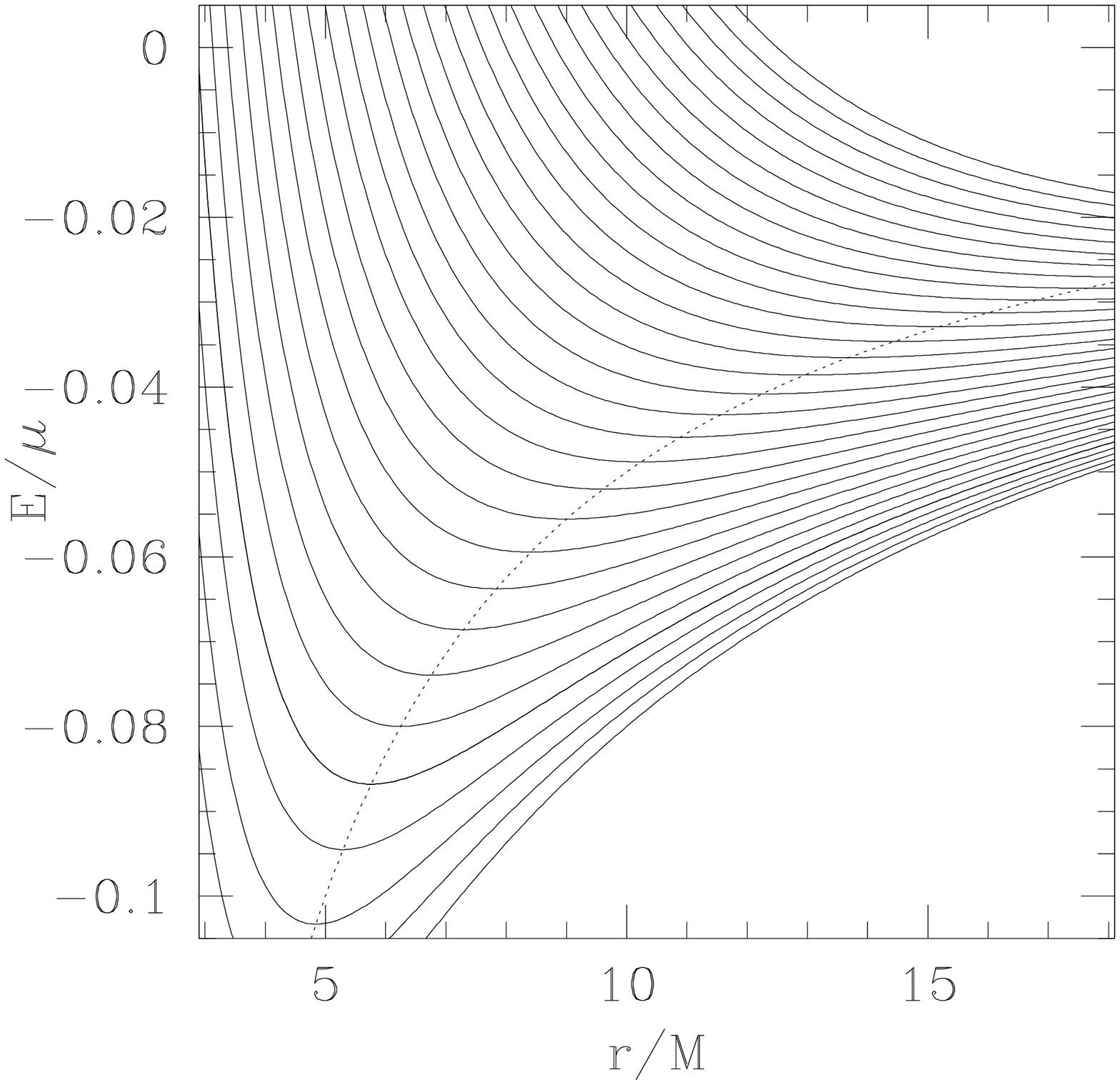,height=3in,width=3in}
	\epsfig{file=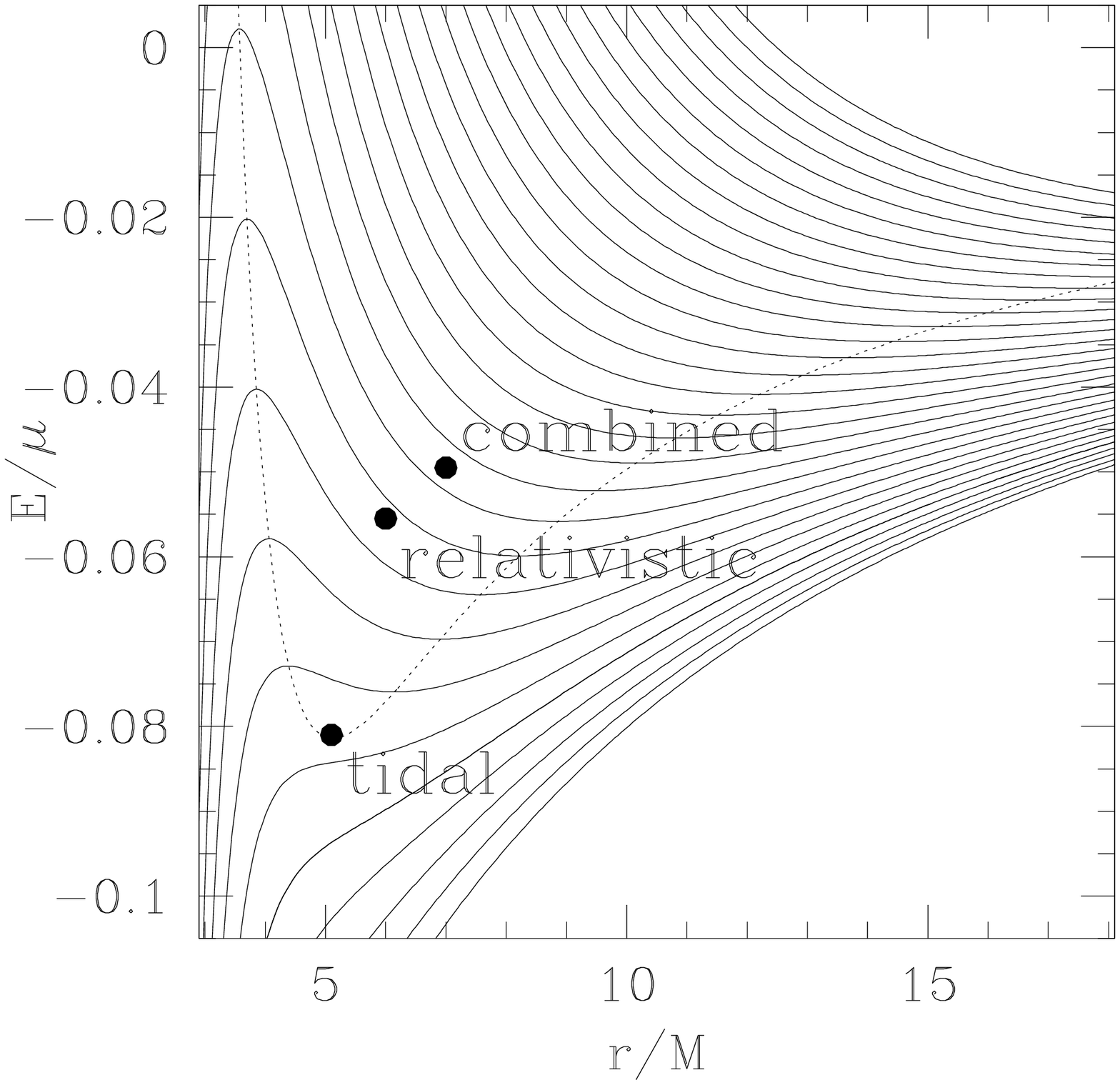,height=3in,width=3in}}
\vspace{10pt}
\caption{Energy as function of separation for a Newtonian point mass
binary (left panel), and a model problem including tidal effects
(right panel, see text).  The solid lines are contours of constant
angular momentum $J$, with the values of $J$ increasing from the
bottom to the top.  The extrema of these contours correspond to
circular orbits (eq.~(\ref{dEdr})).  The dashed line connects the
circular orbits and represents the equilibrium energy $E_{\rm eq}$,
and the dots mark the ISCO for tidal, relativistic and combined 
effects.}
\label{fig1}
\end{figure}

Note that Newtonian point masses can be brought into arbitrarily close
and arbitrarily tightly bound circular orbits.  The orbits are stable
for all separations $r$, indicating that in Newtonian point mass
binaries there is no ISCO.  As we will see, this is related to the
gravitational interaction potential being proportional to $1/r$.

It is also illustrative to construct circular orbits graphically.  In the
left panel of Fig.~\ref{fig1}, the solid lines are contours of the
energy~(\ref{E1}) at constant angular momentum.  The minima of these
contours correspond to circular orbits.  Connecting these yields the
equilibrium energy~(\ref{Eeq}) as a function of separation.

\subsection*{Tidal and relativistic effects}

Compact binaries are extended, relativistic objects, of course, so
that we have to take into account both tidal and relativistic effects.

For irrotational, identical stars, tidal effects can be modeled by
adding a tidal interaction term
\begin{equation} \label{Etidal}
E_{\rm tidal} = - 2 \lambda \frac{\mu M R^{5}}{r^{6}}
\end{equation}
to the energy~(\ref{E1}) (compare \cite{lrs93a,l96,lw96}), where $R$
is the radius of the stars at large separations.  The coefficient
$\lambda$ depends on how easily the stars can be deformed, and hence
on their structure and equation of state (EOS).  For polytropic EOSs
with polytropic index $n$ we have $\lambda = (3/4)\,\kappa_n (1 -
n/5)$, where the coefficient $\kappa_n$ is tabulated, for example, in
\cite{lrs93b}.  For an incompressible fluid, $\lambda = 3/4$.  Each
star induces a quadrupole moment in the companion\footnote{If the
stars are not irrotational, the spin of the individual stars induces
an intrinsic quadrupole moment.  The tidal interaction energy then
scales with a smaller power of $r$, see, e.g.~\cite{lrs93a,lrs94}}
which scales with $1/r^3$, and the two quadrupole moments' interaction
again scales with $1/r^3$, so that $E_{\rm tidal} \sim r^{-6}$.

From eq.~(\ref{dEdr}), we can now find the equilibrium energy
\begin{equation}
E_{\rm eq} = - \frac{1}{2}\,\frac{\mu M}{r} 
	+ 4 \lambda \frac{\mu M R^{5}}{r^{6}}.
\end{equation}
We immediately see that we can no longer construct arbitrarily tightly
bound orbits, and that instead $E_{\rm eq}$ now assumes a minimum for
a finite separation $r$.  It is easy to show that any attractive
interaction potential proportional to $1/r^n$, $n>2$, leads to a
positive contribution to $E_{\rm eq}$, and hence the existence of a
minimum at finite $r$.

In the right panel of Fig.~\ref{fig1} we provide a graphical
representation for incompressible fluid stars with $m/R = 0.2$, where
$m$ is the mass of the individual stars.  For large separations, the
tidal interaction is very small, and the orbits are similar to those
of point masses.  For small separations, however, the tidal
interaction becomes dominant, and decreases the energy of a
configuration with a given separation and angular momentum below that
of the point mass binary.  For large enough angular momentum, the $J =
const$ contours now have a maximum at a small separation in addition
to the minimum at a larger separation, while for small angular momenta
their contours no longer have any extrema.  As a consequence, the
equilibrium energy goes through a minimum.  Outside of this minimum,
the equilibrium energy curve connects minima of the $J = const$
curves, which correspond to {\em stable} circular orbits, while inside
this minimum, the curve connects maxima of $J = const$ curves, which
correspond to {\em unstable} circular orbits.  The minimum of the
equilibrium curve therefore marks the innermost stable circular orbit,
the ISCO\footnote{At this point, a word of warning is in order.
Relativistic binaries emit gravitational radiation, causing them to
slowly spiral towards each other, and they hence do not follow
strictly circular orbits.  The very concept of an innermost stable
{\em circular} orbit is therefore somewhat ill defined.  Also, the
minimum in the equilibrium energy identifies the onset of a {\em
secular} instability, while the onset of {\em dynamical} instability
may be more relevant for the binary inspiral (see, e.g., the
discussion in~\cite{lrs93b} and also~\cite{lrs97}, where it is shown
that the two instabilities coincide in irrotational binaries).
Moreover, it has been suggested that the passage through the ISCO may
proceed quite gradually~\cite{bd00}, so that a precise definition
of the ISCO may be less meaningful than the above turning method
suggests.  Ultimately, dynamical evolution calculations will have to
simulate the approach to the ISCO and to investigate these issues.
For the sake of dealing with a well-defined problem, we will here
identify the ISCO with the minimum of the equilibrium energy.}.

We can similarly mimic relativistic effects by borrowing a relativistic
interaction potential
\begin{equation}
E_{\rm rel} = - \frac{M}{\mu}\,\frac{J^2}{r^3}
\end{equation}
from the result for test particles in Schwarzschild space times.  This
interaction again gives rise to an ISCO.  We mark the location of this
ISCO in the left panel of Fig.~\ref{fig1}, as well as the ISCO when
computed from the combined tidal and relativistic effects.  It is
evident from this model calculation that including relativistic
effects will move the ISCO to larger separation, and correspondingly
smaller values of the orbital frequency $\Omega$.

Obviously, this model problem is very naive, and can at best mimic
qualitative effects.  However, it does illustrate several important
results.  Quite in general, the ISCO arises because of contributions
to the interaction potential which deviate from a simple $1/r$
scaling.  In particular, tidal effects can cause an ISCO even in
purely Newtonian systems.  In general, both tidal and relativistic
effects have to be taken into account for an accurate determination of
the ISCO.  Lastly, the model problem provides a straight-forward
recipe for locating the ISCO: first construct circular orbits and the
corresponding equilibrium energy $E_{\rm eq}$ by locating extrema of
the energy (eq.~(\ref{dEdr})), then identify the ISCO with the minimum
of $E_{\rm eq}$, and last compute the orbital frequency $\Omega$ from
eq.~(\ref{dEdJ}).  We have therefore reduced the problem of accurately
determining the ISCO in black hole or neutron star binaries to an
accurate determination of their energies, which we will address
separately in the following sections.

\section*{The ISCO in binary black holes}

Determining the ISCO in binary black hole systems has been attempted
with two independent approaches: post-Newtonian expansions and numerical
calculations in full general relativity.  We will outline both approaches,
and will then compare the results.

\subsection*{Post-Newtonian Calculations}

A large number of researchers has developed post-Newtonian techniques
to model compact objects in close binaries (an incomplete list
includes~\cite{ww76,dis98,js98,bd99,djs00}; see~\cite{ww96} for
a particularly pedagogical treatment).  Typically, these calculations
start by bringing Einstein's equations into the form
\begin{equation} \label{PN_field}
\Box h^{\alpha\beta} = - 16 \pi \tau^{\alpha\beta}
\equiv - 16 \pi (-g) T^{\alpha\beta} - \Lambda^{\alpha\beta}.
\end{equation}
Here, $\Box$ is the flat space wave operator, $h^{\alpha\beta}$
measures deviations from the flat Minkowski metric, $g$ is the
determinant of the metric, and the source term $\tau^{\alpha\beta}$
contains both the stress-energy tensor $T^{\alpha\beta}$ as well as
non-linear terms in $h^{\alpha\beta}$, which have been absorbed in
$\Lambda^{\alpha\beta}$.  A formal solution to this equations is
the retarded, flat-space Green function
\begin{equation} \label{PN_int}
h^{\alpha\beta}(t,{\bf x}) = 4 \int 
	\frac{\tau^{\alpha\beta}(t',{\bf x}') 
	\delta(t' - t + |{\bf x} - {\bf x}'|)}{|{\bf x} - {\bf x}'|}
	d^4 x'.
\end{equation}
Unfortunately, the source term $\tau^{\alpha\beta}$ depends on the
solution $h^{\alpha\beta}$, so that this formal solution is of little
practical help.  However, it can be used to construct a solution
iteratively.  Starting with a Newtonian point-mass solution, for which
$T^{\alpha\beta}_{\rm Newt}$ is known and for which $h^{\alpha\beta}$
vanishes, we can construct a first iteration from
\begin{equation}
\Box h^{\alpha\beta}_1 = 16 \pi T^{\alpha\beta}_{\rm Newt}.
\end{equation}
Each solution $h^{\alpha\beta}_n$ can be inserted on the right hand
side of eq.~(\ref{PN_int}), which then provides the next iteration
$h^{\alpha\beta}_{n+1}$.  Each iteration provides a correction over
the previous one in the order of
$\epsilon \sim v^2 \sim m/R$.
The iteration therefore yields a post-Newtonian expansion of the 
solution $h^{\alpha\beta}$, from which the energy can be constructed
in the form of a Taylor expansion
\begin{equation} \label{PN_e}
E_{\rm eq} = E_{\rm Newt} + E_1\,\epsilon + E_2\,\epsilon^2 + \ldots .
\end{equation}
Here $E_n\,\epsilon^n$ is the $n$-th order post-Newtonian correction
to the energy.

It turns out, however, that for values close to the ISCO, $\epsilon
\sim 1/6$, this expansion converges extremely slowly.  The reason is
that at a location fairly close to the ISCO, namely at the light
radius with $\epsilon \sim 1/3$, some of the functions involved in the
expansion diverge and have a pole (see~\cite{dis98}).  It is therefore
to be expected that a polynomial expansion of the form~(\ref{PN_e})
will converge only very poorly in the neighborhood of that pole.
Damour, Iyer and Sathyaprakash~\cite{dis98} provided a solution to
this problem by using a resummation technique.  Instead of employing
the polynomial~(\ref{PN_e}), they use the information contained in the
Taylor expansion to construct an expansion in terms of rational
functions.  To illustrate this with an example, assume that we know
the Taylor expansion of an arbitrary function $f(x)$ up to second
order.  We can then construct an expansion in terms of a rational
function
\begin{equation}
f(x) \sim t_0 + t_1\,x + t_2\,x^2 + \ldots 
	\sim \frac{p_0 + p_1\,x + \dots}{1 + q_1\,x + \dots}
\end{equation}
by choosing the coefficients $p_0$, $p_1$ and $q_1$ such that the
value and the first two derivatives of the two expansions match at
$x=0$.  This expansion is known as a ``Pad\'e-approximant'', and has
been shown to greatly improve the convergence of post-Newtonian
expansions at least up to second order.

When $h^{\alpha\beta}_n$ is inserted on the right hand side of
eq.~(\ref{PN_int}), the integrals no longer have compact support, so
that there is no guarantee that they will converge.  Moreover, the use
of point-mass sources gives rise to divergent integrals, which have to
be re-normalized appropriately.  This has been achieved up to second
post-Newtonian order, but at third post-Newtonian order some
ambiguities remain~\cite{js98}.  Damour, Jaranowski and
Sch\"afer~\cite{djs00} have recently shown that these ambiguities can
be expressed by a single dimensionless parameter, $\omega_{\rm
static}$, which is currently unknown.  They compare three different
post-Newtonian approaches (the $e$-method and $j$-method based on
minimizations of the energy and the angular momentum, and an effective
one-body method~\cite{bd99}) and find a quite remarkable result.  For
one particular choice of $\omega_{\rm static}$, these three
independent approaches yield very similar predictions for the angular
momentum, energy, and orbital frequency at the ISCO.  This
self-consistency suggests that the correct values of $\omega_{\rm
static}$ and the ISCO at 3PN order may have been identified.

\subsection*{Numerical Calculations in General Relativity}

A framework for numerically constructing models of binary black holes
has been provided by Arnowitt, Deser and Misner's 3+1 (ADM)
decomposition of Einstein's equations~\cite{adm62} and York's
conformal decomposition~\cite{york}.

The ADM formalism splits Einstein's equations into constraint and
evolution equations.  The gravitational fields, described by the
spatial metric $\gamma_{ij}$ and the extrinsic curvature $K_{ij}$,
have to satisfy the constraint equations on each time slice, while the
evolution equations describe how they evolve from one time-slice to
the next.  For the construction of initial data, the two constraint
equations, the Hamiltonian constraint and the momentum constraint,
have to be solved, which determine only the longitudinal parts of the
gravitational fields.  The transverse parts of the fields, loosely
associated with the gravitational wave degrees of freedom, are
unconstrained by the constraint equations, and have to be chosen
before the constraints can be solved.

Since the binary inspiral outside the ISCO proceeds very slowly, the
gravitational wave content of these spacetimes must be very small.  It
therefore seems reasonable to attempt to minimize the gravitational
wave content by choosing the spatial metric conformally flat,
$\gamma_{ij} = \psi^4 f_{ij}$, where $\psi$ the conformal factor and
$f_{ij}$ a flat metric.  With the further assumption of maximal
slicing, $K \equiv \gamma^{ij} K_{ij} = 0$, the Hamiltonian constraint
reduces to
\begin{equation} \label{Ham}
\hat \nabla^2 \psi = - \frac{1}{8} \psi^{-7} \hat A_{ij} \hat A^{ij},
\end{equation}
while the momentum constraint becomes
\begin{equation} \label{mom}
\hat \nabla_i \hat A^{ij} = 0.
\end{equation}
Here $\hat \nabla$ is the flat space covariant derivative, and $\hat
A^{ij}$ is the trace-free part of the conformally related extrinsic
curvature, $\hat A_{ij} = \psi^2 (K_{ij} - \gamma_{ij} K/3)$.  Quite
remarkably, the momentum constraint~(\ref{mom}) is now a linear
equation and decouples from the Hamiltonian constraint.  Solutions
describing a pair of black holes with arbitrary momenta and spins can
now be constructed analytically by super-imposing two solutions for
single black holes~\cite{by80}.  Given these solutions, the
Hamiltonian constraint~(\ref{Ham}) can then be solved
numerically~\cite{c91,bb97}, which completes the construction of
binary black hole initial data.

To construct binary black hole models in circular orbits and to
determine their ISCO, turning points of the energy of these solutions
have to be located, in complete analogy to the model problem
above~\cite{c94,b00}\footnote{The two approaches of~\cite{c94}
and~\cite{b00} differ in the topology chosen for the binary black
hole:~\cite{c94} adopts the two-sheeted topology of~\cite{c91},
whereas~\cite{b00} assumes the three-sheeted topology of~\cite{bb97}.
Their results agree to within a few percent, indicating that this
choice has little effect on the ISCO.}.  These results have recently
been generalized to binaries in which the individual black holes carry
spin~\cite{ptc00}.

\subsection*{Comparison}

\begin{figure}[t] 
\centerline{\epsfig{file=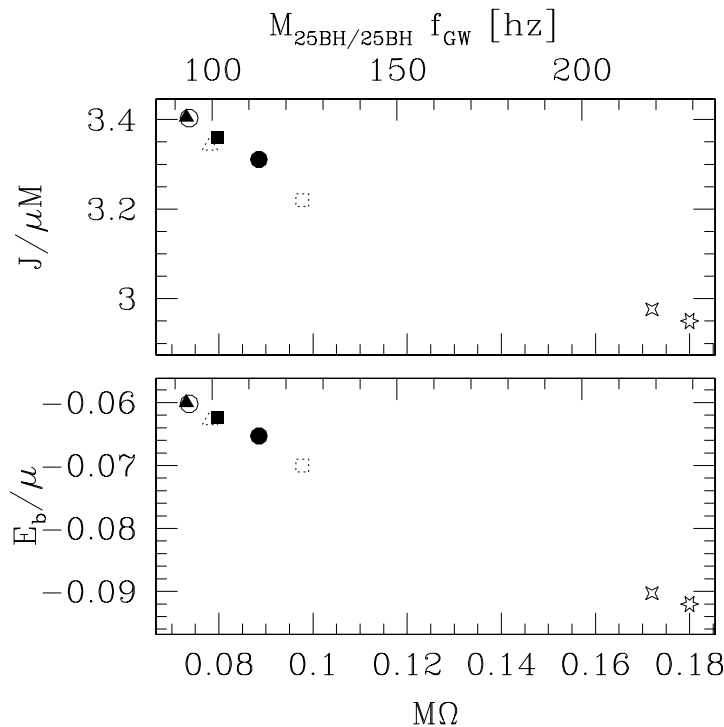,height=4in,width=4in}}
\vspace{10pt}
\caption{Results for the angular momentum, energy and orbital
frequency of a black hole binary at the ISCO from post-Newtonian and
numerical calculations.  The solid triangle, square and circle are the
2PN results of~\protect\cite{djs00} using the effective one-body
method, the $e$-method, and the $j$-method.  The open circle is their
3PN result, with the unknown parameter $\omega_{\rm static}$ chosen
such that all three methods agree.  The four-pointed and six-pointed
star are the numerical results of~\protect\cite{c94}
and~\protect\cite{b00}.  The dashed triangle and square are 2PN
results using the effective one-body and $e$-method together with a
conformal-flatness assumption.  The top label gives the corresponding
gravitational wave frequencies for a binary of two 25 $M_{\odot}$
black holes. }
\label{fig2}
\end{figure}

In Fig.~\ref{fig2}, we summarize results from the post-Newtonian and
numerical calculations.  It is obvious that both the 2PN and 3PN
results differ quite significantly from the numerical results, by more
than a factor of two in the frequency.  This discrepancy is very
unsatisfactory, and should be resolved by re-examining both approaches
and their assumptions.

The PN approach treats the binary stars as point-masses, and hence
neglects tidal effects (even though these could be included, see,
e.g., Appendix F of~\cite{ww96}).  This approximation is probably
fairly poor for binary neutron stars, but may be more adequate for
binary black holes.  Moreover, including tidal effects would probably
move the ISCO to a larger separation and smaller orbital frequency,
and would hence increase the discrepancy between the PN and numerical
results.  However, the convergence properties of the PN expansion and
the lacking renormalization of higher-order expansion coefficients
remain worrisome.

The biggest worry in the numerical calculations is probably the
assumption of conformal flatness.  The effect of this assumption is
evaluated in Appendix B of Ref.~\cite{djs00}, where the authors
determine the ISCO at 2PN order, using their effective one-body and
$e$-method together with a conformal-flatness assumption.  These
results are included as the dashed triangle and square in
Fig.~\ref{fig2}.  Obviously, this assumption changes the results
somewhat, but not nearly enough to explain the discrepancy between the
PN and numerical results.  When evaluated for the effective one-body
method, it even seems like conformal flatness is quite an adequate
approximation.

None of the above more obvious worries seem to be able to completely
explain the differences between the PN and numerical results.  This
suggests that perhaps the two approaches construct sequences which are
not physically identical.  In our discussion of the model problem
above, we have implicitly assumed that the masses $m_1$ and $m_2$ of
the binary stars remain constant during the inspiral.  In general
relativity, however, there is no unique definition of the mass of the
individual black holes in a binary, and it is not clear which mass is
conserved during the inspiral.  It has been conjectured~\cite{b99}
that during the adiabatic inspiral outside of the ISCO the
``irreducible mass''~\cite{c70} of the black hole event horizons is
conserved.  In the numerical calculations of~\cite{c91,b00}, this is
approximated by the irreducible mass of the black hole apparent
horizons\footnote{Locating an event horizon requires knowledge of a
complete spacetime, while an apparent horizon can be located on a
single timeslice.}.  In the PN approaches the masses $m_1$
and $m_2$ of the point sources are kept constant.  It is not obvious
that the two approximations are equivalent, and it is therefore
possible that the two approaches construct physically distinct
sequences.

\section*{The ISCO in binary neutron stars}

The ISCO in binary neutron stars has been computed in numerous
Newtonian~\cite{lrs93b,lrs93a,lrs94,rs,nt97},
PN~\cite{lrs97,lw96,stn97} and relativistic
calculations~\cite{bcsst97,irrotational,use00}.  We will briefly
summarize some of the qualitative findings of these calculations, and
will then discuss some of the more recent relativistic results.

\subsection*{Qualitative discussion}

For binary neutron stars, the location of the ISCO depends on the
physical properties of the individual stars, in particular their EOS
and spin.  Qualitatively, these effects can be illustrated by
re-examining the tidal correction to the energy~(\ref{Etidal}).
Computing $E_{\rm eq}$ from~(\ref{dEdr}) and~(\ref{Etidal}), and then
locating the minimum of $E_{\rm eq}$, shows that the ISCO occurs at a
separation
\begin{equation} \label{risco}
\frac{r_{\rm ISCO}}{R} = ( 48 \lambda )^{1/5}
\end{equation}
and a frequency
\begin{equation} \label{fisco}
m \, \Omega_{\rm ISCO} = \sqrt{\frac{5}{2}} \,
	( 48 \lambda )^{-3/10} \left( \frac{m}{R} \right)^{3/2}.
\end{equation}
Note that $r_{\rm ISCO}$ scales with the stellar radius $R$, and $m
\Omega_{\rm ISCO}$ accordingly with $(m/R)^{3/2}$, where $m/R$ is the
stellar compaction (compare~\cite{lrs94,lw96,use00}).

The coefficient $\lambda$ depends on the EOS.  For polytropic EOSs,
$\lambda$ increases with decreasing polytropic index $n$.
Eq.~(\ref{risco}) therefore implies that $r_{\rm ISCO}/R$ is larger
for stiffer EOSs.  This explains why an ISCO exists only for
stars with sufficiently stiff EOSs; for stars with too soft EOSs the
stars merge before they encounter the ISCO.  The critical value of $n$
for the existence of an ISCO depends on the rotation of the stars and
on relativistic effects, but is generally believed to be fairly close
to $n_{\rm crit} \sim 1$.  Eq.~(\ref{fisco}) also implies that the
ISCO frequency is correspondingly smaller for stiffer EOSs.

In the above discussion we have assumed that the stars are
irrotational, so that $E_{\rm tidal} \sim r^{-6}$.  The effect of
individual rotation can be estimated by allowing $E_{\rm tidal}$ to
scale with a different power of $r$
(compare~\cite{lrs97,lrs93a,lrs94}).

\subsection*{Numerical Calculations in General Relativity}

Constructing relativistic binary neutron stars requires solving the
equations of relativistic hydrodynamics together with the initial
value equations of general relativity.  This problem can be simplified
whenever the fluid flow is stationary, so that the equations of
hydrodynamics reduce to a relativistic Bernoulli
equation\footnote{Note that a solution to the Bernoulli equation is by
construction in equilibrium, and yields the equilibrium energy $E_{\rm
eq}$.  For binary black holes the latter has to be constructed by
finding turning points of the energy $E$.}.

This approach was first adopted by~\cite{bcsst97}, who constructed
corotational models of binary neutron stars in quasi-equilibrium.  The
initial value equations of general relativity were solved with the
assumption that the spatial metric is conformally flat.  For moderate
(but realistic) compactions ($m/R \lesssim 0.2$), for which the
gravitational fields are only moderately strong, the latter assumption
has been shown to be quite adequate (compare~\cite{uue00}).  Results
for various polytropic indices $n$ and stellar compactions $m/R$ can
be found in~\cite{bcsst97}.  For stars of 1.4 $M_{\odot}$, $n=1$ and
$m/R = 0.2$, for example, they find a gravitational wave frequency at
the ISCO of $\Omega_{\rm ISCO}^{\rm GW} = 2 \, \Omega_{\rm ISCO} \sim
1300\,\mbox{Hz}$.

The viscosity in neutron star interiors is not believed to be strong
enough to maintain corotation during binary inspiral~\cite{viscosity}.
It is therefore more realistic to assume the binary to be
irrotational, in which case the relativistic equations of
hydrodynamics can again be reduced to a Bernoulli
equation~\cite{irrotational_formulation}.  Constructing irrotational
binaries is computationally more complicated than constructing
corotational binaries, because one of the boundary conditions has to
be imposed on the surface of the stars.  The location of the latter
changes during the numerical iteration, and is a priori unknown.
Nevertheless, several groups have succeeded in constructing
relativistic, irrotational models of binary neutron
stars~\cite{irrotational}.

Probably the most careful analysis to date of evolutionary sequences
of irrotational binary neutron stars is presented in~\cite{use00}.  As
pointed out earlier~\cite{irrotational}, the authors find that for
moderately soft EOSs ($n \gtrsim 2/3$), a cusp forms at the surface of
the stars before an ISCO is encountered, at which point the numerical
method breaks down.  Most likely, the cusp indicates that a Lagrange
point forms, and that matter starts to overflow towards the companion.
The assumption of a stationary, irrotational fluid flow seems no
longer adequate, and will have to be relaxed for the construction of
closer binaries.  An ISCO appears in these simulations only for fairly
stiff EOSs ($n \lesssim 2/3$).  For these cases, the authors
of~\cite{use00} quite carefully analyze the ISCO for various
compactions $m/R$, and show that it is dominated by the hydrodynamical
effects of the tidal interaction.  Up to moderate compactions ($m/R
\lesssim 0.2$), the relativistic effects can be expressed as PN
corrections to the Newtonian scaling~(\ref{fisco}).

\section*{Summary}

Knowledge of the ISCO may be an important ingredient for the future
detection of gravitational wave signals and their interpretation.  In
this article we illustrate in a simple model problem how turning-point
methods can be used to locate the ISCO (see, however, the
disclaimer in footnote (2)).  We also summarize recent efforts to
determine the ISCO in both black hole and neutron star binary systems.

Currently, the biggest challenge for an accurate determination of the
ISCO in binary black hole systems seems to be the disagreement between
PN and numerical approaches.  For binary neutron stars, we seem to be
lacking a realistic description of the velocity field in close
binaries.  Ultimately, the quasi-equilibrium approaches presented in
this paper should be backed up with fully dynamical simulations
(e.g.~\cite{su00}).

As a final comment, we point out that in this article we have only
discussed binaries containing either black holes or neutron stars.  So
far black hole-neutron star binaries have only been modeled in
Newtonian or PN dynamical simulations~\cite{bhns1} and in ellipsoidal
model calculations~\cite{bhns2}.  Such mixed binaries have yet to be
constructed in a relativistic framework, and their ISCO (or the onset
of tidal disruption) has yet to be determined self-consistently.

\hspace{0.1in}

It is a pleasure to thank Greg Cook, Stuart Shapiro, and Masaru
Shibata for numerous very useful discussions.  The author gratefully
acknowledges support through a Fortner Fellowship.  This work was also
supported by NSF Grant PHY 99-02833 at Illinois.

\end{document}